\documentclass[aps,pra,twocolumn,superscriptaddress,showpacs,floatfix]{revtex4-2}

\usepackage{lmodern}
\usepackage[T1]{fontenc}
\usepackage[utf8]{inputenc}
\usepackage{amsmath}
\usepackage{amsfonts}
\usepackage{amssymb}
\usepackage{graphicx}
\usepackage{braket}
\usepackage{hyperref}
\usepackage{xcolor}
\hypersetup{
    colorlinks=true,
    linkcolor=blue,
    citecolor=blue,
    urlcolor=blue}

\begin{document}
\title{The Dirac sea of phase: Unifying phase paradoxes and Talbot revivals in multimode waveguides}
\author{N. Korneev}
\email[e-mail:\,]{korneev@inaoep.mx}
\affiliation{Instituto Nacional de Astrofísica Óptica y Electrónica, Calle Luis Enrique Erro No. 1\\ Santa María Tonantzintla, Puebla, 72840, Mexico}
\author{I. Ramos-Prieto}
\email[e-mail:\,]{iran@inaoep.mx}
\affiliation{Instituto Nacional de Astrofísica Óptica y Electrónica, Calle Luis Enrique Erro No. 1\\ Santa María Tonantzintla, Puebla, 72840, Mexico}
\author{H. M. Moya-Cessa}
\affiliation{Instituto Nacional de Astrofísica Óptica y Electrónica, Calle Luis Enrique Erro No. 1\\ Santa María Tonantzintla, Puebla, 72840, Mexico}

\date{\today}
\begin{abstract}
The quantum mechanical description of phase remains a fundamental challenge, with theoretical efforts tracing from the early works of London and Dirac to discrete formalisms. In this work, we extend the action-angle formalism to the Helmholtz-Schrödinger equation by introducing a phase-dependent wavefunction $\phi(\theta, t)$ residing in the Hardy space $H^2(\mathbb{D})$. This mathematical structure, defined by functions analytic on the unit disk with square-integrable boundary values, naturally ensures the positivity of the energy spectrum while providing a rigorous framework for wave dynamics in photonic systems. We demonstrate that establishing a self-adjoint phase operator requires extending the Hilbert space to $L^2$, a procedure that necessitates the admission of negative energy states. We interpret these states through an analogy with the Dirac sea, where the existence of antiphase or antiphoton modes provides a conceptual framework for understanding the fundamental limits of phase localization and quantum uncertainty. This formalism is applied to light propagation in multimode waveguides characterized by anharmonic refractive index profiles. By mapping modal dispersion to our phase representation, we show that the deviation of propagation constants from linear spacing governs the spatial evolution of the optical field. This approach offers a clear mechanism for the emergence of periodic self-imaging known as the Talbot effect, the generation of fractional revivals, and the formation of complex fractal interference patterns, providing a robust toolkit for the characterization and design of multimode interference devices.
\end{abstract}
\maketitle
\section{Introduction} \label{sec:intro}
Multimode waveguides serve as essential components in modern integrated photonics and quantum optical circuits. Compared to single-mode systems, multimode structures offer a significantly expanded operational space, enabling high-dimensional quantum information processing, reconfigurable mesh networks, and high-performance sensing platforms \cite{politi2008, obrien2009, wang2020, pelucchi2022}. The utility of these waveguides is centered on the phenomenon of multimode interference, which facilitates the operation of robust beam splitters, couplers, and switches. While these devices are prized for their high tolerance to fabrication variations \cite{soldano1995, moody2022}, fully exploiting their potential requires a deep understanding of internal phase-space dynamics that go beyond standard modal decomposition.

The traditional description of light propagation in such structures relies on the paraxial Helmholtz equation. This framework establishes a direct mathematical isomorphism between the longitudinal propagation distance $z$ and the temporal evolution $t$ of a quantum state. Based on this formal equivalence, the wave propagation in multimode waveguides can be rigorously described by a Schrödinger-type operator, a mapping that we synthesize under the term Helmholtz-Schrödinger equation. This nomenclature highlights the intersection between classical paraxial optics and quantum mechanics, enabling the direct translation of quantum dynamical toolkits to classical wave systems. In the following, we shall use the terms Schrödinger equation and Schrödinger-like equation interchangeably to refer to the evolution of the optical field along the longitudinal coordinate. While this model provides clear representations for canonical coordinates like position and momentum, it has historically faced difficulties when describing their angular conjugates, namely the action $\hat{J}$ and the phase $\hat{\theta}$ \cite{carruthers1968, lynch1995, barnett1986}.

This difficulty mirrors the long-standing problem in quantum mechanics of defining a Hermitian phase operator, a challenge dating back to the insights of London \cite{london1926} and Dirac \cite{dirac1927}. Dirac initially attempted to define a phase through the polar decomposition of the annihilation operator, but this approach encountered mathematical barriers due to the non-unitarity of the resulting operators and the semi-bounded nature of the photon number spectrum. Over many years, various formalisms have been proposed to resolve the phase problem, including the Susskind-Glogower operators \cite{susskind1964}, the Pegg-Barnett discrete approach \cite{pegg1989, pegg1988, Perez_2016, Ramos_2020, Tapia_2026}, and several methods based on quasiprobability distribution functions \cite{hillery1984, vogel1989}.

In this work, we present an alternative approach by formulating a Schrödinger-like description directly in a continuous angle coordinate. Rather than seeking a problematic operator, we define the quantum state in the Bargmann-Fock space and project it into a phase-basis $\ket{\theta}$ residing in the Hardy space $H^2(\mathbb{D})$ \cite{burkholder1971, alpay2022,dubin2000,bohm2011_ijtp}. The Hardy space consists of analytic functions on the open unit disk whose boundary values on the unit circle are square-integrable. This choice of mathematical structure is motivated by the spectral constraint that the energy, corresponding to photon number or mode index, remain non-negative. This constraint is fundamental to quantum optics and photonic systems, where negative photon numbers are unphysical. The resulting framework provides a description of phase as a continuous dynamical variable while automatically preserving the positivity of the energy spectrum through the analytic structure of the underlying function space. In contrast to previous approaches that impose positivity through truncation or regularization, the Hardy space formulation achieves this constraint through the geometric structure of the mathematical domain itself. We also explore the theoretical implications of extending this description to the complete $L^2$ space, leading to a conceptual Dirac sea of phase interpretation \cite{dirac1930}. By viewing the negative frequency modes allowed in $L^2$ as virtual antiphoton states, we establish a conceptual framework for the limits of phase localization and the underlying analytic structure of quantum phase states \cite{bialynicki1976, bohm2013}.

Finally, we apply this framework to interference challenges in multimode waveguides. We show that the interference pattern governing self-imaging and the Talbot effect \cite{talbot1836, rayleigh1881, wen2013, bryngdahl1973} can be derived as an effective dispersive equation in the phase variable. This allows for the prediction of complex phenomena such as fractional revivals \cite{averbukh1989, robinett2004} and the formation of Talbot carpets \cite{berry2001} within the waveguide structure. The mechanism underlying these phenomena exhibits fundamental parallels to the well-studied collapse and revival dynamics observed in Jaynes-Cummings models of cavity quantum electrodynamics \cite{jaynes1963, shore1993, eberly1980, rempe1987}. In both systems, coherent evolution of an initially localized quantum state is interrupted by the presence of incommensurate energy scales: in Jaynes-Cummings systems, the mismatch between atomic transition frequency and cavity mode frequency induces periodic dephasing and rephasing; in multimode waveguides, the deviation of the modal spectrum from linear spacing plays the analogous role. The quadratic term in the modal energy generates a phase-space diffusion that fragments an initially coherent wave packet, yet the periodicity of the phase domain and the discrete nature of the modal spectrum ensure reversibility and periodic reconstruction. This formal correspondence suggests that optical multimode interference devices can serve as classical analogues for studying quantum coherence phenomena previously accessible only through atomic and cavity quantum electrodynamics, while simultaneously providing new insights for the development of high-performance multimode interference devices \cite{ulrich1978, soldano1995}.

The manuscript is organized as follows. In Sec.~\ref{sec:phase_rep}, we develop the phase representation within the Hardy space, establishing the analytic structure and the Dirac sea interpretation. Sec.~\ref{sec:anharmonic} addresses anharmonic waveguides through operator formalism, numerical spectral analysis, and propagation simulations. Sec.~\ref{sec:conclusion} provides general conclusions.

\section{The phase representation} \label{sec:phase_rep}
Establishing a mathematically rigorous framework for the quantum phase operator remains an enduring challenge in quantum optics \cite{london1926, dirac1927}. Unlike classical wave mechanics, where phase is a continuous and unbounded variable, the quantum description of the electromagnetic field is constrained by the strict positivity of the photon number spectrum. This spectral semi-boundedness breaks the canonical conjugacy between action and angle variables, making the definition of a unique and self-adjoint phase operator problematic \cite{carruthers1968, susskind1964}. Historical efforts to resolve this, from the polar decomposition of Dirac \cite{dirac1927} to the finite-dimensional Hilbert space formalism of Pegg and Barnett \cite{pegg1989, pegg1988}, have often relied on truncations or limiting procedures to avoid the non-unitarity of the exponential phase operator \cite{lynch1995, barnett1986}.

In this section, we move beyond these limitations by adopting a formalism based on the analytic structure of the quantum state. Instead of seeking an operator solution, we construct a phase-space representation that naturally resides within the Hardy space $H^2(\mathbb{D})$ of the unit disk \cite{bohm2013, dubin2000,hardy_note}. This framework enforces the positivity of the energy spectrum as a boundary condition, treating the phase states as rigorous generalized distributions within a rigged Hilbert space, also known as a Gelfand triplet \cite{bohm1989, bohm1998}. By using the isomorphism between the physical wavefunction and the boundary values of functions analytic in the unit disk \cite{schleich2001}, we recover a consistent and physically transparent description of phase evolution \cite{huang2023}. The phase representation is constructed by expressing the field both in the standard modal (number) basis and as a boundary-value function on the unit circle. This hybrid viewpoint avoids the need to define a conventional self-adjoint phase operator inside the restricted physical subspace, and instead uses the analytic properties of boundary traces to enforce spectral positivity.

\subsection{Projection onto the phase variable}
Consider a state expanded in the harmonic-oscillator number basis with amplitudes $C_n(t)$ for $n\ge0$. The phase-dependent boundary function is defined by the one-sided Fourier series
\begin{equation}\label{phi_theta}
    \phi(\theta,t)=\frac{1}{\sqrt{2\pi}}\sum_{n=0}^{\infty} C_n(t) e^{in\theta}, \qquad\theta\in[0,2\pi).
\end{equation}
For the harmonic spectrum ($C_n(t)=C_n(0)e^{-in t}$) this representation yields the rigid-rotation solution $\phi(\theta,t)=\phi(\theta-t,0)$. By construction the Fourier expansion in Eq.~\eqref{phi_theta} contains only non-negative frequencies: this is the defining spectral property of functions in the Hardy space $H^2(\mathbb{D})$, i.e. those boundary traces that arise from functions analytic in the open unit disk  \cite{bohm1989, bohm1981,bohm1998}.

The correspondence between the modal number and the angular variable provides a concrete operator mapping: in the phase representation the number operator acts as a first-order differential operator, so that the action–angle duality is implemented by
\begin{equation}
    \hat{n} \,\longrightarrow\, -i\frac{\partial}{\partial\theta}\quad\text{on }\phi(\theta),
\end{equation}
which implies the usual action–angle relation in the boundary representation:
\begin{equation}
    \langle \theta | \hat{n} | \psi \rangle = -i \frac{\partial}{\partial \theta} \phi(\theta),
\end{equation}
and, for the harmonic Hamiltonian $\hat{\mathcal{H}}=\hat{n}$, the phase evolution takes the transport form
\begin{equation}
    i\frac{\partial \phi}{\partial t} = -i\frac{\partial \phi}{\partial \theta},
    \label{eq:transport}
\end{equation}
Higher-order spectral corrections produce higher-order differential terms in the phase equation (see Sec.~\ref{sec:anharmonic}).

Physically this construction complements the spatial representation: the Hermite–Gauss functions $\psi_n(x)$ form an orthonormal basis in $L^2(\mathbb{R})$ and allow expansion of any transverse profile as
\begin{equation}
    \psi(x,t)=\sum_{n=0}^{\infty} C_n(t)\,\psi_n(x).
\end{equation}
By inspecting Eq.~\eqref{phi_theta}, we observe that for a fixed time, $\phi(\theta)$ possesses a Fourier transform restricted to non-negative frequencies $n \ge 0$. This identifies the wavefunction as an element of the Hardy space $H^2(\mathbb{D})$. A profound consequence of this mathematical structure is the intrinsic locking between the real and imaginary parts of the state, $\phi(\theta) = \phi_R(\theta) + i \phi_I(\theta)$. They are inextricably linked by the Hilbert transform, meaning the complete state is encoded solely in its real or imaginary component. This redundancy, first explored in the context of analytic signal theory, highlights the power of phase-space representations constrained by energy positivity \cite{schleich2001}. Specifically, if the real part admits the sine-cosine expansion:
\begin{equation}
    \phi_R(\theta) = a_0 + \sum_{k=1}^{\infty} a_k \cos(k\theta) - \sum_{k=1}^{\infty} b_k \sin(k\theta),
    \label{eq:phi_r_expansion}
\end{equation}
then the imaginary part is necessarily constrained to:
\begin{equation}
    \phi_I(\theta) = a_{I0} + \sum_{k=1}^{\infty} b_k \cos(k\theta) + \sum_{k=1}^{\infty} a_k \sin(k\theta).
    \label{eq:phi_i_expansion}
\end{equation}
Note that the Schrödinger equation Eq. \eqref{eq:transport} remains valid for $\phi_R$ and $\phi_I$ individually as real equations:
\begin{equation}
    \frac{\partial \phi_R}{\partial t} = -\frac{\partial \phi_R}{\partial \theta}, \quad \frac{\partial \phi_I}{\partial t} = -\frac{\partial \phi_I}{\partial \theta}.
    \label{eq:real_transport}
\end{equation}
Since $\phi_I$ is uniquely determined by $\phi_R$, the evolution of the real part alone is physically equivalent to the full quantum mechanical evolution. This coupling resembles the Kramers-Kronig relations in classical electrodynamics, where causality in the time domain manifests as analyticity in the frequency domain. In our case, the semi-boundedness of the modal spectrum manifests as a causal-like analytic constraint in the angular variable.

Projecting the ket $\ket{\psi(t)}$ onto the position basis $\ket{x}$ yields the standard wavefunction $\psi(x,t) = \braket{x | \psi(t)} = \sum_n C_n e^{-int} \psi_n(x)$. This dual perspective demonstrates that while the position-space description is suited for analyzing beam width and centroid motion, the phase-basis description provides a direct window into the coherence and analytic rephasing properties governing interference phenomena like the Talbot effect \cite{leonhardt1995, schleich2001}.

\subsection{Hardy space, analyticity, and the Hilbert transform}
The restriction of the summation in Eq.~\eqref{phi_theta} to non-negative indices $n \ge 0$ is a physical necessity stemming from the positivity of the energy spectrum. Arno Bohm and his collaborators have argued that the standard Hilbert space axiom should be replaced by a Hardy space axiom to account for time-asymmetry and irreversibility in quantum systems \cite{bohm1999, bohm1997}. Mathematically, $\phi(e^{i\theta})$ serves as the boundary value of a function $\Phi(z)$ that is analytic within the open unit disk $|z| < 1$, where $z = r e^{i\theta}$. This complex analytic structure implies that the state is entirely determined by its boundary behavior, leading to the concept of a beginning of time for individual microsystems, as the evolution is governed by a semigroup rather than a full group \cite{bohm2011_ijtp}. According to the Cauchy integral formula: $\Phi(z) = 1/2\pi i \oint \phi(\zeta)/(\zeta - z) d\zeta$.

The analyticity of $\Phi(z)$ provides a powerful tool for analyzing the stability and convergence of phase-dependent observables. Here $\mathbb{D}$ denotes the open unit disk in the complex plane. Functions belonging to the Hardy space $H^2(\mathbb{D})$ are analytic on $\mathbb{D}$ and have square-integrable radial boundary values on the unit circle. The limit $\phi(\theta)=\lim_{r\to1^-}\Phi(re^{i\theta})$ exists for almost every $\theta$ and defines an element of $L^2([0,2\pi])$. These boundary traces are precisely the phase wavefunctions used throughout this work. Any local disturbance in the phase distribution is instantaneously reflected across the entire unit circle due to the global constraints of analyticity, a feature analogous to the rigid behavior of rigid bodies or the non-locality of quantum entanglement. Recent work has extended these results to weighted Bergman spaces, identifying the philophase states as the minimum uncertainty states in the Hardy representation \cite{huang2023}.

\subsection{The Dirac sea of phase and antiphotons}
The historical phase problem in quantum mechanics stems from the fundamental difficulty of defining a self-adjoint phase operator strictly conjugate to a semi-bounded number operator. Within the Hardy space $H^2(\mathbb{D})$, we cannot consider the multiplication by a real, non-constant function of $\theta$ as a valid operator because its Fourier expansion necessarily includes negative frequency terms that map the state out of the physical subspace. To define a consistent angle operator $\hat{\theta}$, we must extend our domain to the full $L^2[0, 2\pi]$ space. This extension is conceptually analogous to Paul Dirac's introduction of the negative energy sea to explain the stability of atoms and predict the existence of antimatter \cite{dirac1930, bohm1999}.

Just as Dirac postulated that the vacuum is an infinite sea of negative-energy electrons, we treat the $L^2$ extension as a Dirac sea of phase. While both the action operator $\hat{J} = -i\partial_\theta$ and the multiplication operator $\hat{\theta}$ are well-defined in $L^2$, the operator $\hat{\theta}$ is only self-adjoint within this larger space. This extension clarifies the non-unitarity of phase-shifting operations: while the Dirac operator $\exp(i\theta)$ is permitted in $H^2(\mathbb{D})$, its inverse $\exp(-i\theta)$ is not, precluding a unitary description within the Hardy subspace alone.

By postulating that the Schrödinger equation in the phase variable is governed by the rigid transport equation in Eq. \eqref{eq:transport}, we recover the classical result $H=J$. The dynamics describe the displacement of the wavefunction as a whole. However, the complete energy spectrum in $L^2$ includes negative values $E_k = k$ for $k \in \mathbb{Z}$. This mirrors the classical mechanical observation where, although physical energies are positive, the action variable $J$ and the corresponding evolution $\partial_t \theta = \partial H / \partial J = 1$ are mathematically permitted for negative $J$. We interpret these negative spectrum states as antiphoton modes. In this hole-theory heuristic, a hole in the negative-energy phase sea manifests as a physical state with positive energy but an inverted phase evolution. These antiphoton modes provide a dynamical basis for the uncertainty principle, as the virtual interaction with the negative energy domain prevents infinite phase localization while ensuring that the photon number remains non-negative \cite{dubin2000}. This perspective bridges early quantum paradoxes with contemporary efforts to construct self-adjoint phase operators, suggesting that the missing phase information in the Hardy space is effectively stored in a virtual background \cite{bohm2013, bialynicki1976}.

The physical utility of this framework becomes apparent when examining how real waveguide systems deviate from ideal harmonic behavior. While the Hardy space formulation provides exact solutions for perfectly parabolic refractive index profiles, practical systems exhibit anharmonic corrections that drive complex dynamical phenomena. These deviations from linearity, governed by the spectral structure established through our phase representation, manifest as quantum revivals and the classical Talbot effect, connecting our foundational theory directly to observable interference patterns \cite{soldano1995, ulrich1978}.

\section{Anharmonic Dynamics and Revivals} \label{sec:anharmonic}
Practical waveguide systems often deviate from ideal parabolic refractive index profiles, leading to anharmonic modal dispersion relations \cite{milburn1986,soldano1995}. In such systems, the propagation constants cannot be linearized; instead, higher-order terms in the spectral expansion become non-negligible, breaking the equidistant energy level spacing characteristic of the harmonic oscillator. These spectral nonlinearities, arising from geometric perturbations, material chromaticity, or intensity-dependent Kerr effects, fundamentally alter the phase evolution of the optical field. Consequently, the simple periodic reconstruction of wave packets is replaced by complex interference phenomena known as quantum revivals, where the interplay of incommensurate modal frequencies generates intricate spatial structures in the field distribution.

The theoretical treatment of these revivals connects directly to the classical Talbot effect, originally observed in near-field diffraction theory \cite{talbot1836, rayleigh1881}. In the quantum domain, this corresponds to the rephasing of wavefunctions governed by quadratic or higher-order dispersive Hamiltonians, a phenomenon characterized in Rydberg atom dynamics \cite{averbukh1989, robinett2004}. In the context of multimode photonics, these interference patterns, manifest as Talbot carpets, serve as sensitive probes of the underlying modal spectrum \cite{wen2013, berry2001}. By extending the present phase-space formalism to include these anharmonic corrections, we provide a rigorous framework for analyzing fractional revivals and self-imaging effects, which are relevant for the design of integrated multimode interference devices and optical signal processing architectures \cite{soldano1995, ulrich1978}.

\color{black}
\subsection{Operator formalism and numerical spectral analysis}
In the framework of paraxial waveguide optics, the quantum mechanical potential $V(x)$ maps directly to the refractive index profile contrast, proportional to $n_0^2 - n^2(x)$, where $n_0$ is the effective core index. An ideal parabolic index profile corresponds to the harmonic oscillator, yielding an equidistant spectrum of propagation constants and ensuring perfect periodic image reconstruction. However, realistic graded-index waveguides invariably exhibit deviations from this ideal quadratic form due to fabrication tolerances, material diffusion, or nonlinear Kerr effects. We model these deviations as an anharmonic perturbation to the effective Hamiltonian, which we express in terms of the position operator $\hat{x}$ and its conjugate momentum $\hat{p}$, using natural units where $\omega=m=\hbar = 1$:
\begin{equation}
    \hat{x} = \frac{1}{\sqrt{2}}(\hat{a} + \hat{a}^\dagger), \quad \hat{p} = \frac{i}{\sqrt{2}}(\hat{a}^\dagger - \hat{a}).
\end{equation}
The Hamiltonian of the system implies a nonlinear evolution given by:
\begin{equation}\label{H}
\begin{split}
    \hat{\mathcal{H}} &=\frac{1}{2}\left(\hat{p}^2+\hat{x}^2\right)+\lambda \hat{x}^4,\\
     &=\hat{a}^\dagger \hat{a} + \frac{1}{2} + \lambda \left[ \frac{1}{\sqrt{2}}(\hat{a} + \hat{a}^\dagger) \right]^4.
\end{split}
\end{equation}
Expanding the quartic term reveals the structure of the couplings. The operator $(\hat{a} + \hat{a}^\dagger)^4$ contains terms such as $(\hat{a}^\dagger)^2 \hat{a}^2$, which represents the photon-number-dependent energy shift characteristic of the Kerr effect in quantum optics \cite{milburn1986}. Since the potential $V(x)$ is symmetric, parity is conserved. The quartic interaction connects the Fock state $\ket{n}$ only to states with the same parity, specifically $\ket{n}, \ket{n\pm 2}$, and $\ket{n\pm 4}$. This selection rule results in a pentadiagonal Hamiltonian matrix in the number basis, which is sparse but contains the off-diagonal elements that mix the modes.

In the harmonic limit where $\lambda\to 0$, the Hamiltonian reduces to that of the simple harmonic oscillator, characterized by an equidistant spectrum $\mathcal{E}_k = k+1/2$. For any finite anharmonicity $\lambda \neq 0$, this symmetry is broken. To solve for the anharmonic eigenvalues and eigenstates, we solve the time-independent Schrödinger equation:
\begin{equation} \label{eq:eigenvalue_problem}
    \hat{\mathcal{H}}\ket{\varphi_k} = \mathcal{E}_k \ket{\varphi_k}.
\end{equation}
Since the analytic solution for the quartic oscillator is not expressible in terms of elementary functions, we employ a numerical diagonalization approach. We expand the physical eigenstates $\ket{\varphi_k}$ in the complete, orthonormal basis of the harmonic oscillator Fock states $\{\ket{n}\}$:
\begin{equation} \label{eq:eigenvector_expansion}
\begin{split}
    \ket{\varphi_k} &= \sum_{n} \braket{n|\varphi_k} \ket{n} \\
    &= \sum_{n} c_n^{(k)} \ket{n},
\end{split}
\end{equation}
where $c_n^{(k)}$ are the mixing coefficients. The numerical implementation involves truncating the Hilbert space to a finite dimension $N_{max}$, chosen to be large enough so that the highest-energy eigenstate of interest has negligible support on the boundary states. By diagonalizing the resulting matrix, we obtain the numerically exact energy spectrum $\mathcal{E}_k$ and the eigenvectors $\ket{\varphi_k}$.

\begin{figure}
    \centering
    \includegraphics[width=\linewidth]{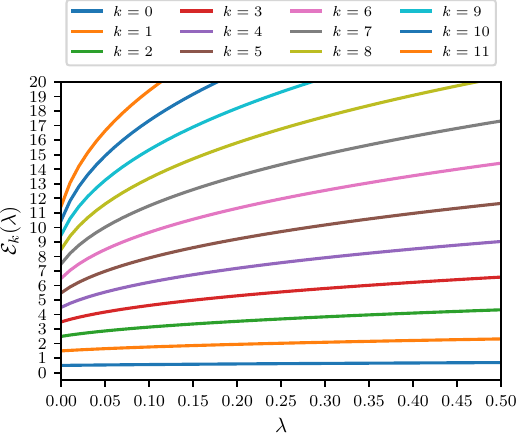}
    \caption{Spectral evolution of the anharmonic oscillator. The plot displays the first twelve eigenvalues $\mathcal{E}_k$ as a function of the anharmonicity parameter $\lambda$. The deviation from the equidistant harmonic ladder increases with both coupling strength $\lambda$ and mode index $k$, indicating the onset of strong nonlinear dispersion required for revivals.}
    \label{fig_1}
\end{figure}
\begin{figure}
    \centering
    \includegraphics[width=\linewidth]{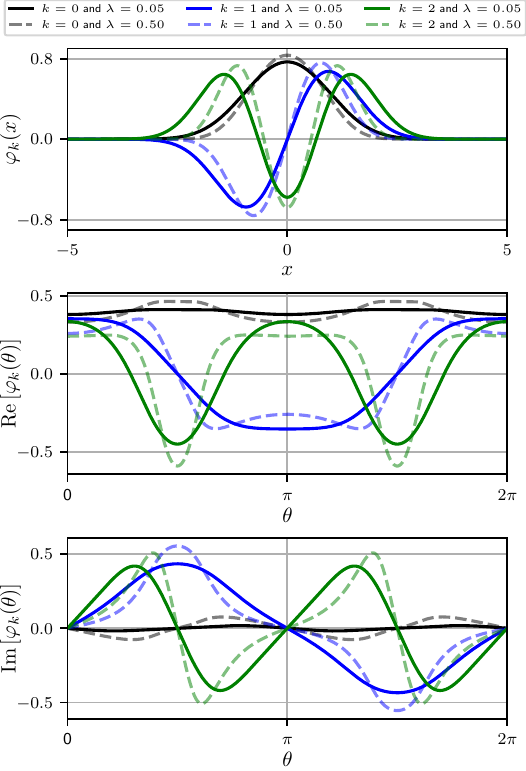}
    \caption{Eigenfunctions of the anharmonic oscillator in dual configuration and phase-space representations. The upper panels show the spatial wavefunctions $\varphi_k(x)$ for two distinct anharmonicity strengths, $\lambda=0.05$ and $\lambda=0.50$, where the steeper quartic potential leads to increased confinement and a characteristic deformation of the nodal structure compared to the ideal harmonic case. The lower panels display the corresponding eigenfunctions projected onto the phase basis, $\varphi_k(\theta)$, separated into their real ($\text{Re}[\varphi_k(\theta)]$) and imaginary ($\text{Im}[\varphi_k(\theta)]$) components. This angular representation reveals the underlying phase topology and localization properties governed by the Hardy space $H^2(\mathbb{D})$ structure, highlighting how the anharmonicity distorts the phase-space symmetry and encodes the nonlinear frequency shifts that drive the Talbot-revival dynamics.}
    \label{fig_2}
\end{figure}

Figure \ref{fig_1} illustrates the behavior of the first twelve eigenvalues of the anharmonic oscillator as a function of the anharmonicity parameter $\lambda$. As $\lambda$ increases, the energy levels progressively diverge from the linear, equidistant spacing characteristic of the harmonic oscillator. This spectral deformation is non-uniform; higher-energy states probe the steeper part of the quartic potential more deeply, leading to a larger perturbative shift. This corresponds to the classical property of anharmonic oscillators where the oscillation frequency depends on the amplitude. This energy-dependent frequency shift is the mechanism behind the dephasing and rephasing dynamics observed in multimode waveguide propagation. Once the spectral decomposition is achieved, the reconstruction of the optical modes in both configuration and phase space is straightforward. The spatial wavefunction $\varphi_k(x)$ is obtained by projecting onto the position basis:
\begin{equation} \label{eq:spatial_proj}
    \varphi_k(x) = \braket{x|\varphi_k} = \sum_{n}c_n^{(k)} \psi_n(x),
\end{equation}
where $\psi_n(x)$ are the standard Hermite-Gauss functions. While valuable, the spatial representation often obscures the phase structure of high-order modes. Therefore, we also compute the phase distribution $\varphi_k(\theta)$ by projecting onto the phase basis:
\begin{equation} \label{eq:phase_proj}
    \varphi_k(\theta) = \frac{1}{\sqrt{2\pi}} \sum_{n} c_n^{(k)} e^{in\theta}.
\end{equation}
This dual projection capability allows for a comprehensive characterization of the anharmonic modes. In Fig. \ref{fig_2}, the spatial wavefunctions $\varphi_k(x)$ are shown, followed by the real and imaginary parts of the eigenfunctions in theta, $\varphi_k(\theta)$. While the spatial representation demonstrates confinement due to the steeper potential, the phase representation provides the distribution of the eigenfunctions in the angular domain, which is fundamental for describing phase evolution. The characterization of these eigenvalues is critical for predicting imaging performance. In paraxial optics, the interference between modes is governed by the phase differences accumulated during propagation. A perfectly linear spectrum leads to perfect self-imaging at the Talbot length. The presence of anharmonicity introduces phase aberrations that distort image formation \cite{soldano1995}. This numerical spectral analysis allows us to quantify the spectral curvature responsible for fractional revivals and the formation of high-contrast Talbot carpets, providing a design tool for multimode interference devices beyond the standard parabolic approximation \cite{ulrich1978}.

\subsection{Waveguide propagation}
\begin{figure*}
    \centering
    \includegraphics[width=\linewidth]{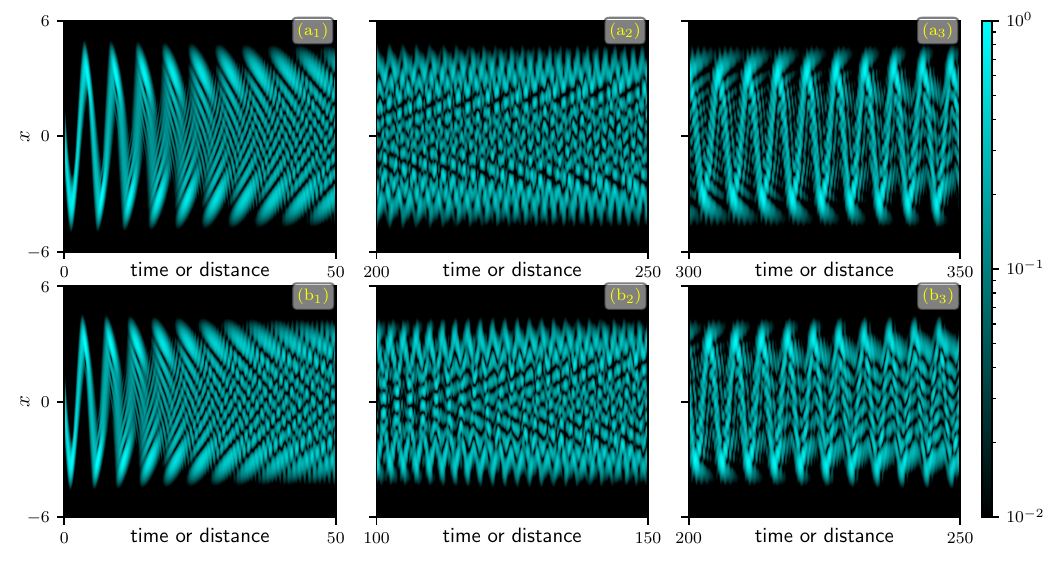}
    \caption{Spatiotemporal evolution of the transverse probability density $|\psi(x,t)|^2$ in an anharmonic waveguide with quartic potential. The density maps illustrate the longitudinal propagation ($t=z$) for two coupling strengths: $(\mathrm{a}_j)$ $\lambda=0.01$ and $(\mathrm{b}_j)$ $\lambda=0.02$, where the index $j=1,2,3$ denotes selected temporal windows corresponding to distinct dynamical regimes. The initial configuration is a coherent state with $\alpha=4i$. Higher anharmonicity is shown to accelerate the collapse-revival cycle, inducing faster modal dephasing and a reduction in the fundamental revival period. These spatial carpets visualize the transition from near-field coherence to complex interference and eventual self-imaging, encoding the underlying modal dispersion of the system.}
    \label{psi_xt}
\end{figure*}
Beyond the static characterization of eigenvalues, the true predictive power of our formalism is revealed in the longitudinal evolution of the optical field along the $t=z$ axis. By applying the unitary propagator $\hat{U}(t) = \exp(-i\hat{\mathcal{H}}t)$ to a prescribed input configuration, we simulate how spectral detuning and nonlinear dispersion drive the spatial and phase-space dynamics. Unlike perturbative approaches that often mask the emergence of revivals, our use of the numerically exact anharmonic basis $\{\ket{\varphi_k}\}$ provides a rigorous path toward understanding how the cumulative phase aberrations determine the fate of the beam.

\begin{figure*}
    \centering
    \includegraphics[width=\linewidth]{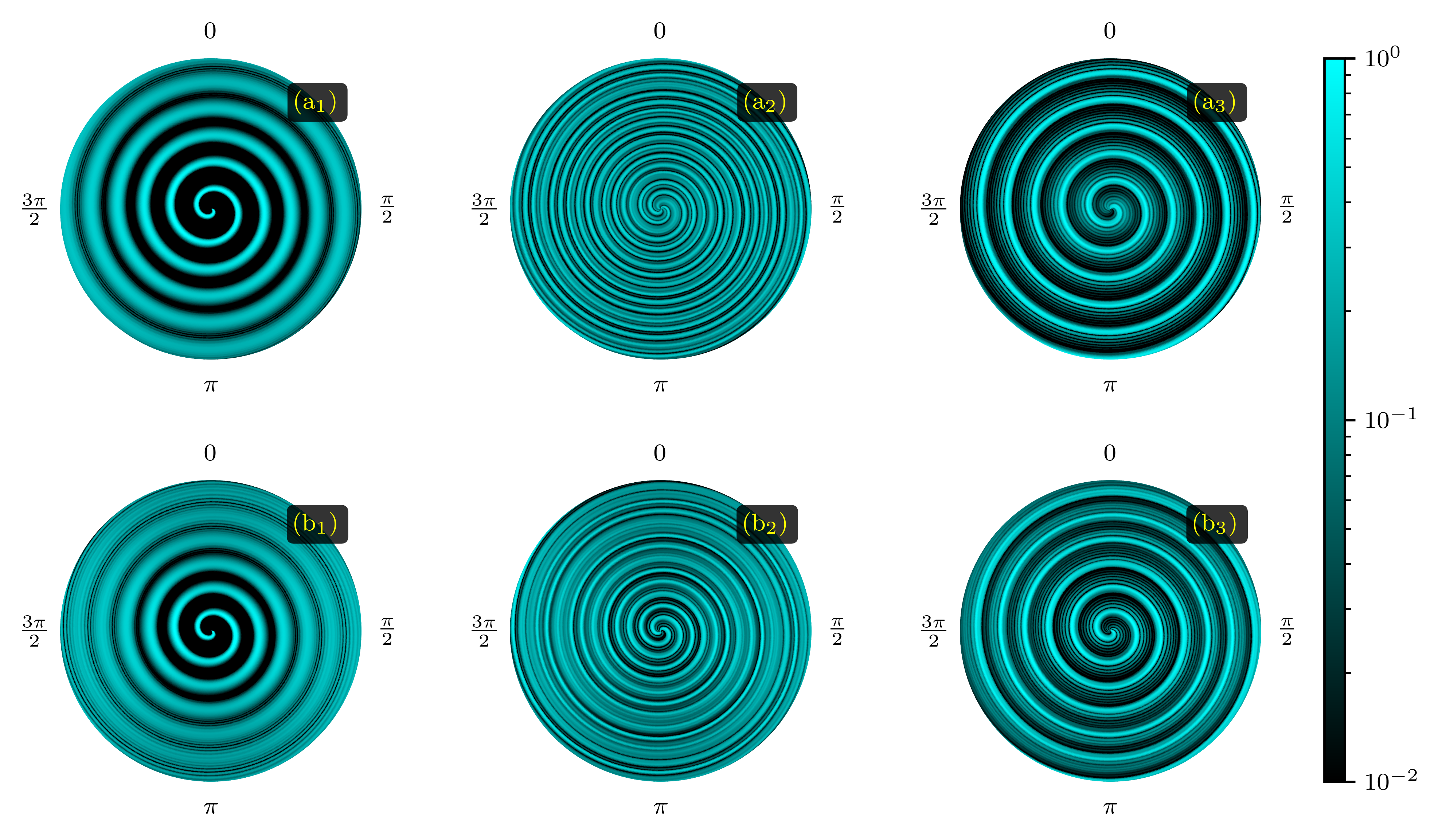}
    \caption{Spatiotemporal evolution of the phase probability density $|\phi(\theta,t)|^2$ within a polar phase-space representation. The density is mapped into a polar coordinate system $(t, \theta)$, where the radial distance represents the propagation axis $t=z$ and the angular coordinate corresponds to the phase $\theta \in [0, 2\pi)$. Panels correspond to coupling strengths $(\mathrm{a}_j)$ $\lambda=0.01$ and $(\mathrm{b}_j)$ $\lambda=0.02$, initialized with a coherent state $\alpha=4i$. This visualization exposes the analytic structure of the state in the Hardy space $H^2(\mathbb{D})$, revealing the formation of fractal Talbot carpets and substructures at fractional revival times. The interplay between the linear rotation and quadratic de-alignment driven by the spectrum $\mathcal{E}_k$ results in the shredding and subsequent periodic reconstruction of the phase distribution.}
    \label{psi_thetat}
\end{figure*}

To model the input field excitation at the waveguide entrance, we consider a general initial state $\ket{\psi(0)}$ expanded in the Fock basis:
\begin{equation}
    \ket{\psi(0)}=\sum_n d_n(0)\ket{n}.
\end{equation}
Standard integrated photonic injections typically utilize coherent states, which serve as semiclassical wavepackets that probe the anharmonic potential. For a coherent state $\ket{\alpha}$, the amplitudes $d_n(0) = e^{-|\alpha|^2/2}\alpha^n/\sqrt{n!}$ follow a Poissonian distribution, where the parameter $|\alpha|^2$ scales with the input power. Utilizing the spectral decomposition and the completeness of the anharmonic basis, the unitary evolution of the state vector is captured by the expansion:
\begin{equation}
\begin{split}
    \ket{\psi(t)}&=\sum_{n=0}^{\infty}d_n(0) e^{-i\hat{\mathcal{H}}t}\ket{n},\\
    &=\sum_{n,k=0}^{\infty}d_n(0) e^{-i\hat{\mathcal{H}}t}\ket{\varphi_k}\braket{\varphi_k|n},\\
    &=\sum_{n,k=0}^{\infty} d_n(0)c_n^{(k)} e^{-i\mathcal{E}_k t}\ket{\varphi_k}.
\end{split}
\end{equation}
In this framework, the matching coefficients $c_n^{(k)} = \braket{\varphi_k|n}$ quantify the projection of the ideal harmonic states onto the actual eigenstates of the quartic potential. This projection fundamentally encodes the squeezing of the modal structure, where the anharmonicity introduces a detuning dependent on the mode index $k$. To track the transformation of the beam across dual domains, we project the evolved state onto the position and phase basis:
\begin{equation}
    \psi(x,t) = \sum_{n,k=0}^{\infty} d_n(0)c_n^{(k)} e^{-i\mathcal{E}_k t}\varphi_k(x), 
\end{equation}
\begin{equation}
    \phi(\theta,t) = \sum_{n,k=0}^{\infty} d_n(0)c_n^{(k)} e^{-i\mathcal{E}_k t}\varphi_k(\theta).
\end{equation}
The spatiotemporal intensity distributions, presented in Figs.~\ref{psi_xt} and \ref{psi_thetat}, offer a visual mapping of the competition between coherence and dispersion. In the spatial intensity carpet (Fig.~\ref{psi_xt}), we observe how the localized coherent peak is progressively fragmented by incommensurate modal phase velocities. This dispersion-induced dephasing manifests as a decay in the field contrast, yet the discrete nature of the spectrum ensures that the information is not lost but instead stored in the complex interference fringes. At integer multiples of the revival length, the phases realign modulo $2\pi$, reconstructing the input profile. We distinguish three fundamental regimes: $(\mathrm{a_1})-(\mathrm{b_1})$ near-field coherent propagation where the beam preserves its Gaussian-like integrity, $(\mathrm{a_2})-(\mathrm{b_2})$ a mid-field fractal regime characterized by high-density interference and Talbot carpets, and $(\mathrm{a_3})-(\mathrm{b_3})$ the self-imaging window where the spectral components rephase to restore the field density.

The phase-space evolution $|\phi(\theta,t)|^2$ (Fig.~\ref{psi_thetat}) provides a more fundamental window into these dynamics, exposing the underlying Hardy space structure. By mapping the propagation onto a polar coordinate system $(t, \theta)$, we visualize the state's evolution within the unit circle, where the radial axis corresponds to time. In this representation, the linear part of the spectrum drives a rigid rotational transport, while the quadratic anharmonicity induces a shredding of the phase-space density. Unlike spatial distributions, which are constrained by the physical waveguide boundaries, the phase-domain representation is constrained by the analyticity of the state in $H^2(\mathbb{D})$. This geometric requirement ensures that the phase information remains globally coupled across the unit circle, allowing for the emergence of the fractal substructures visible at fractional revival times. The persistence of the Talbot carpet even under strong anharmonicity suggests that the Hardy space formulation possesses an intrinsic stability, providing a robust signature of the modal spectrum and the fundamental limits of quantum phase localization.

\subsection{Talbot revivals and dispersive phase dynamics}
To connect the operator formalism with the phase representation, we use the action-angle correspondence $\braket{\theta|\hat{n}|\psi}=-i\frac{\partial}{\partial\theta}\phi(\theta)$. For weak anharmonicity $\lambda \ll 1$, the Hamiltonian in Eq.~\eqref{H} admits the expansion:
\begin{equation}\label{H_approx}
    \hat{\mathcal{H}}\approx\left(1+\frac{3}{2}\lambda\right)\hat{n}+\frac{3}{2}\lambda\hat{n}^2+\left(\frac{1}{2}+\frac{3}{4}\lambda\right).
\end{equation}
Projecting this onto the phase basis and retaining terms up to second order in the differential operator yields the effective dispersive evolution equation:
\begin{equation}\label{eq:dispersive_phi}
    \begin{split}
    i\frac{\partial \phi}{\partial t} &= -i\left(1+\frac{3}{2}\lambda\right)\frac{\partial \phi}{\partial \theta} + \frac{3}{2}\lambda\frac{\partial^2 \phi}{\partial \theta^2},\\
    &=-a_1i\frac{\partial \phi}{\partial \theta} +a_2\frac{\partial^2 \phi}{\partial \theta^2},
    \end{split}
\end{equation}
where $a_1=1+\frac{3}{2}\lambda$ and $a_2=\frac{3}{2}\lambda$. This equation governs the phase-space dynamics in the anharmonic regime and is the analog of the paraxial wave equation in diffractive optics. The first-order derivative term represents group velocity dispersion and induces rigid rotation. The second-order term introduces phase diffusion, the feature of anharmonicity that breaks the periodicity of the harmonic oscillator. The periodic boundary condition $\phi(\theta+2\pi,t)=\phi(\theta,t)$ ensures that this diffusion is reversible, leading to quantum revivals.

The physical mechanism underlying the Talbot effect in multimode waveguides is clear. The anharmonic spectrum $\mathcal{E}_k$ breaks equidistant modal spacing, causing differential phase accumulation. An initially localized coherent state dephases as modes acquire incommensurate phases proportional to their spectral detuning. This manifests as fragmentation of the transverse beam profile and the spreading of the phase distribution $|\phi(\theta,t)|^2$. At specific propagation distances known as Talbot lengths $T_{\text{Talbot}} = 2\pi m/a_2$, the accumulated phases realign modulo $2\pi$, and the field distribution is reconstituted. This cyclic collapse and revival of coherence is the optical manifestation of the Talbot effect \cite{talbot1836, rayleigh1881, berry2001, wen2013, bryngdahl1973}.

The dispersive coefficient $a_2 = \frac{3}{2}\lambda$ quantifies the deviation from linear modal dispersion. In the harmonic limit where $a_2 = 0$, Eq.~\eqref{eq:dispersive_phi} reduces to a rigid transport equation describing perfect self-imaging. For finite $a_2$, the quadratic phase modulation causes modal components to dephase, with the revival time scaling as $T_{\text{rev}} \sim 2\pi/a_2 \propto 1/\lambda$. This scaling law is the optical analog of the collapse-revival dynamics observed in cavity quantum electrodynamics \cite{jaynes1963, shore1993, eberly1980, rempe1987}, where the quadratic energy dependence induces periodic rephasing. Fractional revivals arise from the mathematical properties of the modal frequencies, generating the intricate Talbot carpet structure visible in Figs.~\ref{psi_xt} and \ref{psi_thetat}.

\begin{figure}
    \centering
    \includegraphics[width=\linewidth]{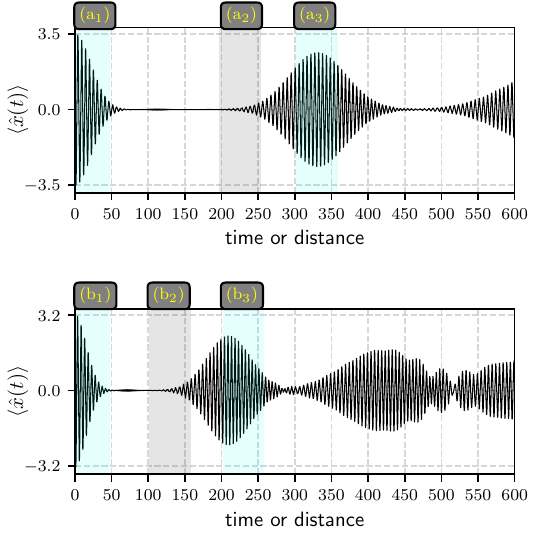}
    \caption{Temporal evolution of the position expectation value $\langle \hat{x}(t) \rangle$ in the anharmonic waveguide, demonstrating Talbot-type collapse and revival dynamics. The state undergoes periodic dephasing and rephasing governed by quadratic spectral dispersion characterized by $a_2=\frac{3}{2}\lambda$. Upper panel: $\lambda=0.01$ ($a_2=0.015$) exhibits a longer revival period $T_{\text{rev}} \approx [300-400]$ with high-fidelity reconstruction. Lower panel: $\lambda=0.02$ ($a_2=0.03$) accelerates the cycle, reducing $T_{\text{rev}}$ by a factor of two while introducing degradation in subsequent revivals. The reconstitution of $\langle \hat{x}(t) \rangle$ at integer multiples of $T_{\text{rev}}$ is the temporal signature of the Talbot self-imaging effect.}
    \label{fig_4}
\end{figure}

To quantify the revival dynamics, we compute the expectation value of the position operator $\langle \hat{x}(t) \rangle = \braket{\psi(t)|\hat{x}|\psi(t)}$. Optically, this observable represents the centroid or center of mass of the beam distribution, corresponding in the classical limit to the average trajectory of the wavepacket as it oscillates within the waveguide. As such, $\langle \hat{x}(t) \rangle$ serves as a sensitive probe of the coherence properties of the optical field. Figure~\ref{fig_4} displays $\langle \hat{x}(t) \rangle$ for an initial coherent state $\alpha=4i$ and anharmonicity parameters $\lambda=0.01$ and $\lambda=0.02$. The temporal evolution shows the signature of Talbot revivals: initial coherent oscillation followed by a collapse phase where the centroid motion is dampened as $\langle \hat{x}(t) \rangle$ decays toward zero. At the revival time $T_{\text{rev}} \approx 2\pi/a_2$, the expectation value reconstitutes, and the beam centroid regains its initial oscillatory amplitude. This collapse-revival cycle repeats periodically, with quality degrading due to higher-order anharmonic terms that break exact commensurability.

The physical interpretation in the phase representation is instructive. The anharmonicity-induced quadratic dispersion causes the initially localized phase distribution $|\phi(\theta,0)|^2$ to spread and fragment in the angular domain, similar to free-particle diffusion. However, the compactness of the phase domain $\theta \in [0,2\pi)$ and the discrete modal spectrum produce periodic refocusing. At the revival time, all modal phases return to their initial configurations modulo $2\pi$, restoring both the phase distribution $|\phi(\theta,T_{\text{rev}})|^2$ and the spatial intensity profile $|\psi(x,T_{\text{rev}})|^2$. This mechanism is the essence of the Talbot effect and provides a route to engineering self-imaging devices by controlling the refractive index profile \cite{soldano1995, ulrich1978}.

\section{Discussion and Conclusion} \label{sec:conclusion}
In this work, we have established a unified theoretical framework that bridges the foundational paradoxes of quantum phase with the observable spatiotemporal dynamics of multimode waveguides. By formulating the Helmholtz-Schrödinger evolution within the Hardy space $H^2(\mathbb{D})$ on the unit disk, we have demonstrated that the long-standing challenges identified by London \cite{london1926} and Dirac \cite{dirac1927}—namely the non-unitarity of the exponential phase operator and the semi-boundedness of the energy spectrum—find a natural resolution in the analytic structure of the underlying function space. In this representation, the positivity of the photon number (or mode index) is not an external constraint but an intrinsic geometric property of the Hardy space, ensuring that every phase wavefunction $\phi(\theta, t)$ represents a physically admissible state without the need for truncation or limiting procedures.

A central conceptual pillar of our work is the introduction of the Dirac sea of phase. By extending the physical Hardy subspace to the full $L^2$ domain, we have established a self-adjoint phase operator whose spectral properties necessitate the admission of negative-energy modes. We interpret these states as virtual antiphoton modes that constitute a stable background for the quantum vacuum in the phase domain. This hole-theory heuristic not only clarifies the mathematical necessity of an extended Hilbert space for self-adjointness but also provides a dynamical basis for the uncertainty principle: the virtual interactions with these antiphoton states prevent the infinite localization of phase, thereby protecting the positivity of the number operator. This unification provides a rigorous answer to the first part of our title, transforming the phase paradoxes into a consistent picture of analytic field evolution.

Furthermore, we have shown that this foundational formalism provides a powerful predictive toolkit for multimode photonics. By mapping the anharmonic modal dispersion of a quartic waveguide onto our phase representation, we derived an effective dispersive evolution equation that governs the formation of Talbot revivals. Our results demonstrate that the deviation from linear modal spacing—the optical analog of the collapse-revival dynamics in cavity quantum electrodynamics \cite{rempe1987}—manifests as a structured shredding and subsequent rephasing of the phase distribution. The resulting fractal Talbot carpets (Fig. \ref{psi_thetat}) are not merely interference patterns but direct visualizations of the global analytic constraints imposed by the Hardy space. The analyticity ensures that phase de-alignment is globally coordinated across the unit circle, facilitating the periodic reconstruction of the input signal at the Talbot length.

From a practical perspective, this framework offers a robust design strategy for integrated photonic devices. Our explicit parametrization of the revival dynamics through the dispersive coefficients $a_1$ and $a_2$ allows for the engineering of high-performance multimode interference couplers and broadband sensors that operate beyond the standard parabolic approximation. The identification of philophase states as minimum uncertainty wavepackets suggests new limits for phase-sensitive imaging and optical metrology \cite{huang2023}. Fundamentally, the Dirac sea interpretation opens new avenues for exploring the limits of light localization, particularly in non-Hermitian and parity-time symmetric waveguides where loss and gain can be used to modulate the shredding rate and control the revival fidelity \cite{longhi2018, ozdemir2019}.

In conclusion, the unification of the Dirac sea concept with Talbot revival dynamics provides a consistent narrative for wave dynamics in photonic circuits. By treating phase as a continuous dynamical variable residing in a space that preserves both causality and spectral positivity, we have transformed a foundational problem of quantum mechanics into a practical tool for integrated optics. This bridge between the abstract Dirac sea of phase and the tangible Talbot revivals establishes a rigorous foundation for future developments in structured light, quantum information processing, and the next generation of multimode interference devices.


\begin{thebibliography}{50}
\bibitem [{har()}]{hardy_note}%
  \BibitemOpen
  \href@noop {} {}\bibinfo {note} {The letter $\mathbb{D}$ is added to denote the open unit disk in the complex plane as the domain for $H^p$ functions}\BibitemShut {NoStop}%
\end{thebibliography}
%

\end{document}